\documentclass{aa} 

\usepackage{graphicx} 
\usepackage{txfonts} 
\usepackage{natbib} 
 
\newcommand{\Mpc}{$h^{-1}$\thinspace Mpc}

\def\apj{ApJ} 
\def\apjl{ApJL} 
\def\apjs{ApJS} 
\def\aj{AJ} 
\def\aap{A\&A} 
\def\mnras{MNRAS}

\usepackage{amstext}

\begin{document}    
 
\title{Evolution of superclusters and supercluster cocoons in
  various cosmologies} 
 
\author{J. Einasto\inst{1,2,3} 
\and  G. H\"utsi\inst{4} 
\and  I. Suhhonenko\inst{1} 
\and  L. J. Liivam\"agi\inst{1}
\and M. Einasto\inst{1}   
}
\institute{Tartu Observatory, University of Tartu, EE-61602 T\~oravere, Estonia 
\and  
ICRANet, Piazza della Repubblica 10, 65122 Pescara, Italy 
\and 
Estonian Academy of Sciences, 10130 Tallinn, Estonia
\and
National Institute of Chemical Physics and Biophysics, 
Tallinn 10143, Estonia  
} 
\date{ Received 06/05/2020; accepted 05/01/2021}  
 
\authorrunning{Einasto et al.} 
 
\titlerunning{Evolution of superclusters} 
 
\offprints{Jaan Einasto, e-mail: jaan.einasto@ut.ee} 
 
\abstract {} {We investigate the evolution of superclusters and
  supercluster cocoons (basins of attraction), and the effect of
  cosmological parameters on the evolution.}
{We performed numerical simulations of the evolution of the cosmic web
  for different cosmological models: the $\Lambda$ cold dark matter  (LCDM) model with a
  conventional value of the dark energy (DE) density, the open model
  OCDM with no DE, the standard SCDM model with no DE, and the
  hyper-DE HCDM model with an enhanced DE density value. We find
  ensembles of superclusters of these models for five evolutionary
  stages, corresponding to the present epoch $z=0$, and to redshifts
  $z=1,~3,~10,\text{and }~30$. We used the diameters of the largest superclusters and
  the number of superclusters as percolation functions to describe
  the properties of the ensemble of superclusters in the cosmic web.  We
  analysed the size and mass distribution of superclusters in models
  and in real samples based on the Sloan Digital Sky Survey (SDSS). }
{In all models, the numbers and volumes of supercluster cocoons are
  independent of the cosmological epochs.  The supercluster masses increase
  with time and the geometrical sizes in comoving coordinates decrease
  with time for all models.  The LCDM, OCDM, and HCDM models have almost
  similar percolation parameters.  This suggests that the essential
  parameter, which defines the evolution of superclusters, is the
  matter density.  The DE density affects the growth of the
  amplitude of density perturbations and the growth of masses of
  superclusters, but significantly weaker.  The HCDM model
  has the highest speed of the growth of the density
  fluctuation amplitude and the largest growth of supercluster masses during
  the evolution.  The geometrical diameters and the numbers of HCDM
  superclusters at high threshold densities are larger than for the LCDM
  and OCDM superclusters.  The SCDM model has about twice as many
  superclusters as other models, and the SCDM superclusters have
  smaller 
  diameters and lower masses. }
{  We find that supercluster embryos form at very early
    cosmological epochs and that the evolution of superclusters
    occurs mainly inside their cocoons.  The evolution of
    superclusters and their cocoons as derived from density fields
    agress well with the evolution found from velocity
    fields.  }


\maketitle

\section{Introduction} 
 
According to the currently accepted cosmological paradigm, the
evolution of the structure of the universe started from a tiny
perturbation of the primordial medium.  The evolution of perturbations
is affected by the physical content of the matter-energy medium and
by physical processes, from inflation to matter and radiation
equilibrium and beyond. The basic constituents of the matter-energy medium
are dark matter (DM), dark energy (DE), and baryonic matter.  For given
initial density perturbations, the evolution depends on the fractional
density of DM and  DE, which is expressed in units of the total
matter-energy density, $\Omega_{DM}$, and $\Omega_{\Lambda}$.

The structure of the cosmic web depends on initial density
fluctuations and on various gravitational and physical processes during
the evolution.  Differences due to cosmological matter-energy density
parameters affect the structure of the cosmic web on various
scales and the time evolution of the web.  The differences in the
structure of the cosmic web between cold dark matter (CDM) and hot
dark matter (HDM) models are well known. They affect the structure
of the cosmic web on all scales.  The differences in the structure of
models with variable cosmological parameters in the CDM model were
studied by \citet{Angulo:2010aa}.

The differences in cosmological parameters affect the structure of
superclusters of galaxies, which are the largest structures of the cosmic web.
Until recently, superclusters were selected using the matter density
field (\citet{Einasto:2007tg}, \citet{Luparello:2011fr},
  \citet{Liivamagi:2012}). \citet{Tully:2014} suggested defining
superclusters on the basis of their dynamical effect on the cosmic
environment, basins of attraction (BoA), as the volumes containing all
points whose velocity flow lines converge on a given attractor.  By
this definition, BoAs mean both superclusters and their surrounding
low-density regions. To keep the traditional definition of superclusters
as connected high-density regions of  the cosmic web,
\citet{Einasto:2019fk} proposed to call the basins of attraction 
``cocoons''.  Superclusters are high-density regions of their cocoons.

The goal of the present paper is twofold: to investigate the evolution
of superclusters and their cocoons, and to  study the effect of
cosmological parameters on the properties and the evolution of superclusters
and their cocoons.
We accept the CDM paradigm and study deviations from the standard CDM
picture that are due to variations of the DM and DE content.  In this approach we
ignore deviations from the concordance $\Lambda$CDM model
\citep{Bahcall:1999aa}.  These deviations are well known, see for
example \citet{Frieman:2008aa} and \citet{Di-Valentino:2020ac, 
  Di-Valentino:2020aa, Di-Valentino:2020ab}. We assume that these
deviations are smaller than the deviations that are due to variations in DM and DE
content, and that they can be ignored in the present study.  We perform
numerical simulations of the evolution of the cosmic web in a box of
size 1024~\Mpc, using four different sets of cosmological density
parameters.  In three sets we use a constant DM content and vary the DE
content from zero (the open OCDM model), the conventional
$\Lambda$CDM model, and a model with enhanced DE content HCDM (not to
be confused with hot-cold DM models, also denoted HCDM).  The first of
these models has an open cosmology, the second a flat 
cosmology, and the third a closed cosmology.  The fourth model is the
classical standard CDM  (SCDM) model of critical density with no DE;
it also has a flat cosmology.  All models have identical initial phases.
This facilitates determining differences between models.

We use the extended percolation analysis by
\citet{Einasto:2018aa} to describe the large-scale geometry of the
cosmic web.  In this method, superclusters are searched using density
fields that are smoothed with an 8~\Mpc\ kernel.  We find superclusters of
these models for five epochs, corresponding to the present epoch
$z=0$, and to redshifts $z=1,~3,~10,\text{and}~30$, and compare the properties of
the model superclusters with the properties of observed superclusters.  We
also derive the size and mass distributions of superclusters.  The model size
and mass distributions are compared with the
size and luminosity distributions of the observed superclusters in the main galaxy survey of the Sloan Digital
Sky Survey (SDSS).  In calculating the density 
field, we used all DM particles of the simulations. The present study is a
follow-up of the study by \citet{Einasto:2018aa, Einasto:2019fk} of
the evolution of $\Lambda$CDM superclusters, using a broader set of
cosmological parameters.  The evolution of supercluster BoAs was
investigated  by \citet{Dupuy:2020aa} using velocity fields.  This allows us to
compare the evolution of superclusters and their BoAs  in more
detail.

The paper is organised as follows. In the next section we describe
the calculation of the density fields of the simulated and observed samples and
the methods we used to find superclusters and their parameters.  In section 3 we analyse
the evolution of superclusters as described by percolation
functions. In section 4 we discuss the evolution of superclusters in
various cosmological models and compare our results, based on the
analysis of density fields, with the study of the supercluster
evolution using velocity fields.  The last section summarises and 
concludes the paper.

\section{Data} 
 
To find superclusters, we determined the supercluster definition
method and the basic parameters of the method.  We used the density
field method. We defined superclusters as the largest non-percolating
high-density regions of the cosmic web, which host galaxies and
clusters of galaxies, connected by filaments.  Based on our
experience, we used for the supercluster search the matter density field
(the luminosity density field for the SDSS sample), calculated with the
$B_3$ spline with a kernel size of 8~\Mpc.  The determination of the second
parameter of the supercluster search, the threshold density, is
discussed below.

\subsection{Simulation of the cosmic web }
 
\begin{figure}[ht] 
\centering 
\resizebox{0.48\textwidth}{!}{\includegraphics*{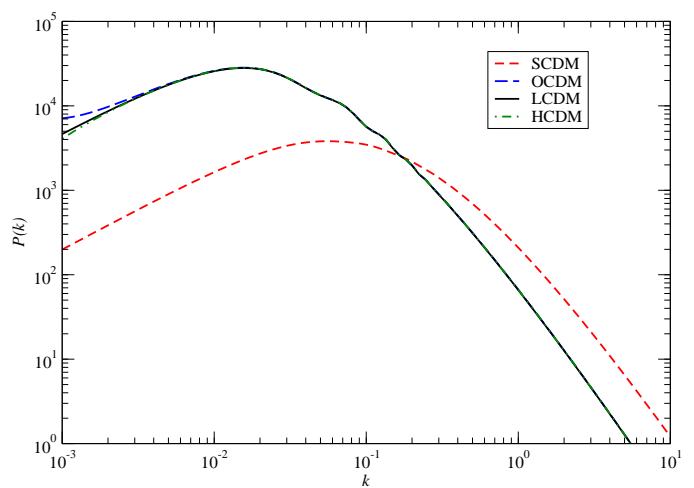}} 
\caption{Linear power spectra of LCDM, HCDM,  OCDM, and SCDM models  at the
  present epoch.}
\label{fig:spectr} 
\end{figure}

To determine the effect of cosmological parameters on the formation of
superclusters, we performed four simulations with different values of
the density parameters.  In the concordance  $\Lambda$CDM model
\citep{Bahcall:1999aa}, we
accepted parameters $\Omega_{\mathrm{m}} = 0.286$,
$\Omega_{\Lambda} = 0.714$.  In the classical standard SCDM model, we
used parameters \citep{Davis:1985} $\Omega_{\mathrm{m}} = 1.000$ and
$\Omega_{\Lambda} = 0$.  In the open OCDM model, we used
$\Omega_m=0.286$, $\Omega_\Lambda=0$.  In the fourth
hyper-DE model HCDM, we assumed
parameters $\Omega_{\mathrm{m}} = 0.286$, and a higher DE density,
$\Omega_{\Lambda} = 0.914$.  This model is not to be confused with the
hot-cold DM model, which is often denoted HCDM, but is not used in
this paper.  In all models we accepted the dimensionless Hubble
constant $h = 0.6932$, and the amplitude of the linear power spectrum
on the scale 8~\Mpc, $\sigma_8 =0.825$. The model parameters are given in
Table~\ref{Tab1}.  The linear power spectra of the density perturbation at the
present epoch are shown in Fig.~\ref{fig:spectr}.

All models have the same realisation, so that the role of different values
of the cosmological parameters can be easily compared. The initial density
fluctuation spectra were generated using the COSMICS code by
\citet{Bertschinger:1995}.  To generate the initial data, we used the
baryonic matter density $\Omega_{\mathrm{b}}= 0.044$
\citep{Tegmark:2004}. Calculations were performed with the GADGET-2
code by \citet{Springel:2005}.  Particle positions and density fields
were extracted for seven epochs between redshifts $z = 30, \dots 0$. We
selected large-scale over-density regions at five cosmological epochs,
corresponding to redshifts $z=0$, $z=1$, $z=3$, $z=10,$ and $z=30$.
The resolution of all simulations was $N_{\mathrm{part}}=N_{\mathrm{cells}}=512^3$, the
size of the simulation boxes was $L_0=1024$~\Mpc, the volume of
the simulation box was $V_0=1024^3$~(\Mpc)$^3$, and the size of the
simulation cell was 2~\Mpc.  This box size is sufficient to see the
role of large-scale density perturbations in the evolution of the
cosmic web.  Using conventional terminology, we call relatively
isolated high-density regions of the cosmic web clusters
\citep{Stauffer:1979aa}.  These clusters are candidates in the search
for  superclusters of
galaxies.  Superclusters have characteristic lengths of up to
$\approx 100$~\Mpc\ \citep{Liivamagi:2012}. As shown by
\citet{Klypin:2019aa}, larger simulation boxes are not needed to
understand the main properties of the cosmic web.  We designate the
simulation with the conventional cosmological parameters LCDM.z,
the standard model with high matter content SCDM.z, the model with
enhanced DE content HCDM.z, and the open model as OCDM.z,
where the index $z$ shows the redshift.

{\scriptsize 
\begin{table}[ht] 
\caption{Model parameters } 
\begin{tabular}{llrrrrc}
\hline  \hline
  Model    & $L_0$&$\Omega_{\mathrm{m}}$ & $\Omega_{\Lambda}$
  & $\Omega_{\mathrm{tot}}$   &$\sigma_8$ & $m_p$\\  
\hline  
(1)&(2)&(3)&(4)&(5)&(6)&(7) \\ 
\hline  \\
LCDM   & 1024&0.286 &  0.714 &    1.000  & 0.825  &  6.355e+11 \\ 
HCDM  & 1024& 0.286 & 0.914  &   1.200  & 0.825  & 6.355e+11 \\
OCDM   & 1024&0.286 &  0.000 &    0.286  & 0.825  &  6.355e+11 \\ 
SCDM  & 1024&1.000 &  0.000  &    1.000 & 0.825 & 2.220e+12 
\label{Tab1}                         
\end{tabular} 
\tablefoot{
Table columns are
\noindent (1): the model name,
\noindent (2): the model side length in \Mpc,
\noindent (3): the $\Omega_{\mathrm{m}}$ -- model matter density,
\noindent (4): the  $\Omega_{\Lambda}$  -- model DE density,
\noindent (5): the $\Omega_{\mathrm{tot}}$ -- model total density,
\noindent (6): the $\sigma_8$ -- amplitude of density perturbations, and
\noindent (7): the  $m_p$ -- particle mass in solar units.
}
\end{table} 
} 

\subsection{SDSS data} 

The density field method allows us to use flux-limited galaxy samples,
and to take galaxies statistically into account that are too faint to be
included in the flux-limited samples, as applied for example by
\citet{Einasto:2003a, Einasto:2007tg} and \citet{Liivamagi:2012} to select
galaxy superclusters. We used the SDSS Data Release 8 (DR8)
\citep{Aihara:2011aa} and the galaxy group catalogue by
\citet{Tempel:2012aa} to calculate the luminosity density field.  In
calculating the luminosity density field, we  took the selection
effects in flux-limited samples into account 
(\citet{Tempel:2009sp}, \citet{Tago:2010ij}). In calculating the
luminosity density field, we selected galaxies  within the apparent
{\em r} magnitude interval $12.5 \le m_r \le 17.77$
\citep{Liivamagi:2012}.  In the nearby region, relatively faint
galaxies were included in the sample, but in more distant regions, only the
brightest galaxies are seen.  To take this into account, we calculated
a distance-dependent weight factor, $W_L(d)$, following
\citet{Einasto:2018aa}.  The weight factor $W_L(d)$ increases to
$\approx 8$ at the far end of the sample; for a more detailed
description of the calculation of the luminosity density field and
the corrections we used, see \citet{Liivamagi:2012}.  The algorithm we used to find
superclusters is described below.  The volume of the SDSS main galaxy
sample is $509^3$~(\Mpc)$^3$ \citep{Liivamagi:2012}.

\subsection{Calculation of the density field}

{ We calculated the smoothed density field using a $B_3$ spline
  \citep[see][]{Martinez:2002fu}.  The $B_3$ spline function is very
  similar to a Gaussian kernel, but has no extended wings. It is different
  from zero only in the interval $x\in[-2,2]$.  The $B_3$ kernel of
  radius $R_B=1$~\Mpc\ corresponds to a Gaussian kernel with
  dispersion $R_G = 0.6$~\Mpc\ \citep{Tempel:2014uq}.  To calculate
  the high-resolution density field, we used a kernel of a scale
  equal to the cell size of the simulation, $L_0/N_{\mathrm{grid}}$,
  where $L_0$ is the size of the simulation box, and
  $N_{\mathrm{grid}}$ is the number of grid elements in one
  coordinate.  The smoothing with index $i$ has a smoothing radius
  $r_i= L_0/N_{\mathrm{grid}} \times 2^i$. The effective scale of
  the smoothing is equal to $2\times r_i$. The non-smoothed density field
  of our models with cell size 2~\Mpc\ corresponds to a kernel
  $R_B =2$~\Mpc. We applied a smoothing with a kernel
  of radius 8~\Mpc, which corresponds to the index 2, and a Gaussian
  kernel $R_G =4.8$~\Mpc.}

{ As shown by \citet{Einasto:2020ab}, smoothing with a $B_3$ kernel
  yields density fields that are partly distorted by the
  insufficient resolution of our models for low-density regions and
  in high-density tails of the density field.  This limits the
  application of the $B_3$ spline methods in high- and low-density
  regions, but does not affect the main results of the analysis. In
  previous studies of superclusters, various methods were used to determine a
  smoothed density field: \citet{Einasto:2007tg} and
  \citet{Luparello:2011fr} used an Epanechnikov kernel  to calculate
  density fields, and \citet{Liivamagi:2012} and  \citet{Einasto:2019fk}
  applied $B_3$ kernel. These studies showed that the basic properties of
  superclusters are almost independent of the density-smoothing
  method. 
}

We calculated for each model the variance of the density contrast,
\begin{equation}
\sigma^2 = 1 /N_{\mathrm{cells}} \sum{(D(\mathbf{x})-1)^2},
\label{disp}
\end{equation}
where $D(\mathbf{x})$ is the density in mean density units at location
$\mathbf{x}$, and summing was performed over all cells of the density field.
 The density field was calculated and its percolation
  functions were determined using the density $D$ as argument.  In the presentation of
  the results we apply percolation functions using as arguments the density
  threshold, reduced to the unit value of the standard deviation  of the density
  contrast:
\begin{equation}
  x = (D_t - 1)/\sigma.
  \label{sigma2}
\end{equation}

\subsection{Finding superclusters}
 
The compilation of the supercluster catalogue consisted of several
steps: calculating the density field,
finding over-density regions as potential superclusters in the density
field, calculating the parameters of potential superclusters, and
finding the supercluster with the largest volume for a given density
threshold.  In this way, we chose the proper threshold
density to compile the actual supercluster catalogue.

We scanned the density field in the range of threshold densities from
$D_t=0.1$ to $D_t=10$ in mean density units.  We used a linear step of
densities, $\Delta D_t = 0.1$, to find over-density
regions. This range covers all densities of practical interest because
in low-density regions the minimum density is $\approx 0.1$, and the
density threshold to find conventional superclusters is
$D_t \approx 5$ \citep{Liivamagi:2012}.  We marked all cells with
density values equal to or higher than the threshold $D_t$ as filled
regions and all cells below this threshold as empty regions.

Inside the first loop, we made another loop over all filled cells to
find neighbours among filled cells. Two cells of the same type were
considered as neighbours (friends) and members of the cluster when
they have a common sidewall. As is traditional in the percolation analysis,
over-density regions are called clusters \citep{Stauffer:1979aa}.
Every cell can have six cells at most as neighbours.  Members of
clusters are selected using a friends-of-friends (FoF) algorithm: the
friend of my friend is my friend.  To exclude very small systems, only
systems with fitness diameters of at least 20~\Mpc\ were added to the
list of over-density regions. These are clusters.

The next step was calculating the cluster parameters.  We
calculated the following parameters: centre coordinates
$x_c, y_c,\text{and } z_c$; cluster diameters (lengths) along the coordinate
axes $\Delta x,~ \Delta y,\text{and }~ \Delta z$; geometrical diameters
(lengths)
$L_{g} = \sqrt{(\Delta x)^2 + (\Delta y)^2 + (\Delta z)^2}$; the fitness
diameters (lengths) $L_f$ discussed in the next subsection;
the geometrical volumes $V_{g}$, which are defined as the volume in space where the
density is equal to or greater than the threshold density $D_t$; and the total
masses $\cal{L}$, which are the mass (luminosity) inside the density contour
$D_t$ of the cluster in units of the mean density of the sample. We
also calculated the total volume of over-density regions, equal to the sum
of volumes of all clusters, $V_C = \sum V_g$, and the respective total
filling factor, $F_f = N_f/N_{\mathrm{cells}}= V_C/V_0$.

During the cluster search we find the cluster with the largest volume
for the given threshold density. We stored in a separate file for each
threshold density the number of clusters found, $N$, and the main data on
the largest cluster: the geometrical diameter $L_{g}$, the fitness
diameter $L_f$, the volume $V_g$, and the total mass (luminosity for
SDSS samples) of the largest cluster.  Diameters are found in \Mpc,
volumes in cubic \Mpc, and the total masses and luminosities in units of the
average cell mass and luminosity of the sample.  These parameters as
functions of the density threshold $D_t$ are called percolation
functions.  They are needed to select the proper threshold density to
compile the actual supercluster catalogue and to characterise general
geometrical properties of superclusters in the cosmic web; for details
of the percolation method, see \citet{Einasto:2018aa}.  In total we
have for every model and evolutionary stage 100 cluster catalogues
(over-density regions) as potential supercluster catalogues. Each
catalogue contains up to 14 thousand clusters
with all cluster parameters mentioned above, depending on the model. These catalogues were
used to find distributions of diameters and cluster masses.

\begin{figure*}[ht]
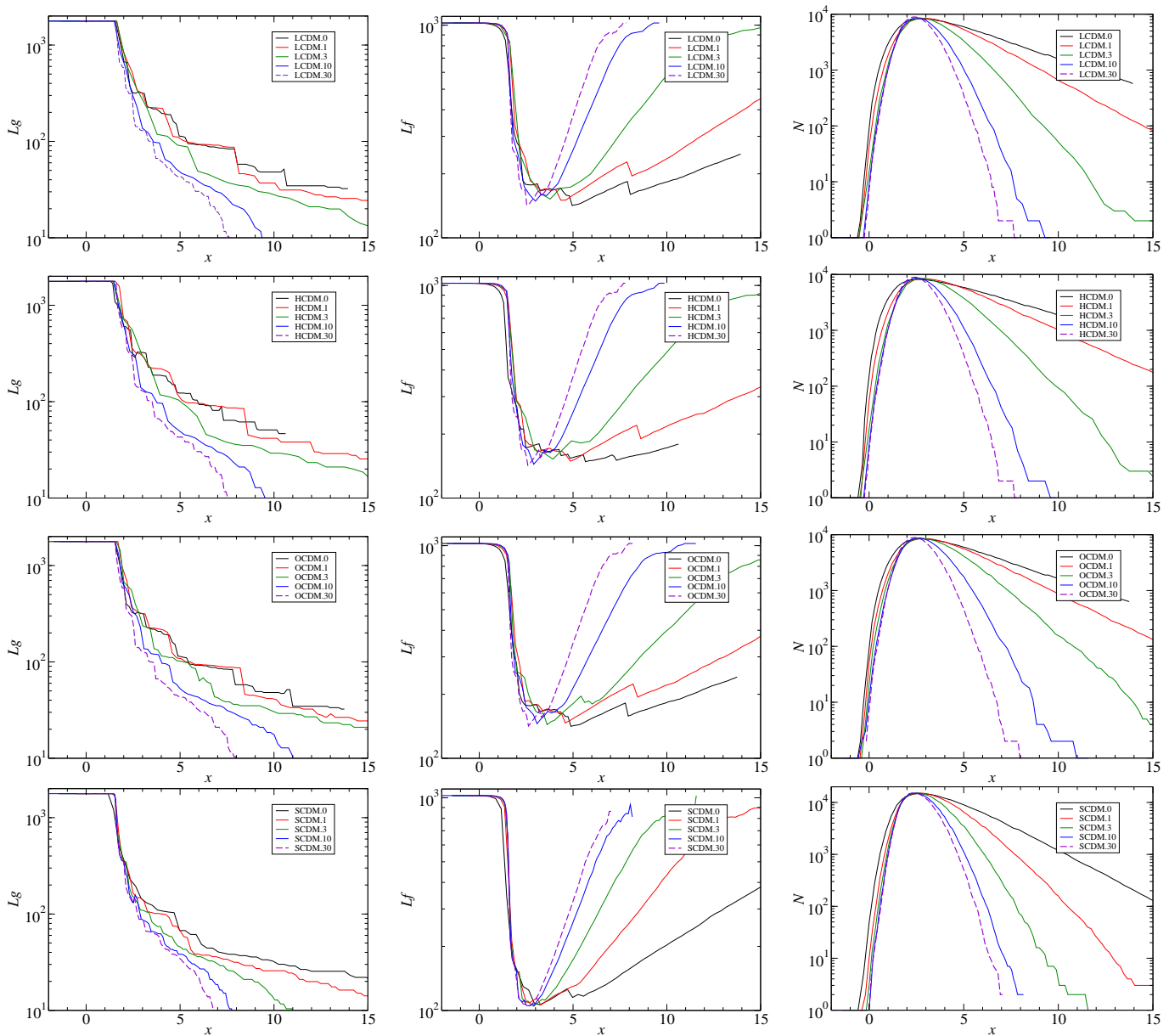
 
\centering 
\hspace{2mm}
 \resizebox{0.31\textwidth}{!}{\includegraphics*{L1024_Lgeom_sigma.eps}}
\hspace{2mm} 
\resizebox{0.31\textwidth}{!}{\includegraphics*{L1024_Ldyn_sigma.eps}}
\hspace{2mm}
\resizebox{0.31\textwidth}{!}{\includegraphics*{L1024_N_sigma.eps}}\\
  \hspace{2mm}
 \resizebox{0.31\textwidth}{!}{\includegraphics*{HCDM_Lgeom_sigma.eps}}
  \hspace{2mm}
  \resizebox{0.31\textwidth}{!}{\includegraphics*{HCDM_Ldyn_sigma.eps}}
 \hspace{2mm}
 \resizebox{0.31\textwidth}{!}{\includegraphics*{HCDM_N_sigma.eps}}\\
  \hspace{2mm}
 \resizebox{0.31\textwidth}{!}{\includegraphics*{OCDM_Lgeom_sigma.eps}}
  \hspace{2mm}
  \resizebox{0.31\textwidth}{!}{\includegraphics*{OCDM_Ldyn_sigma.eps}}
\hspace{2mm}  
  \resizebox{0.31\textwidth}{!}{\includegraphics*{OCDM_N_sigma.eps}}\\
\hspace{2mm} 
  \resizebox{0.31\textwidth}{!}{\includegraphics*{SCDM_Lgeom_sigma.eps}}
  \hspace{2mm}
  \resizebox{0.31\textwidth}{!}{\includegraphics*{SCDM_Ldyn_sigma.eps}}
 \hspace{2mm}
 \resizebox{0.31\textwidth}{!}{\includegraphics*{SCDM_N_sigma.eps}}
  \caption{Percolation functions of models. The {\em left} panels show the
    geometrical length function,  the {\em central} panels the fitness
    length function, and the {\em right} panels the number function.  We used the reduced threshold
    density $x= (D_t - 1)/\sigma$  as
    arguments of  percolation functions.  Diameters are given in \Mpc.
    The panels from {\em top} to {\em
      bottom} show the LCDM, HCDM, OCDM, and SCDM models. }
\label{fig:evol} 
\end{figure*}

{\scriptsize 
\begin{table*}[ht] 
\caption{Model parameters and SDSS superclusters.} 
\begin{tabular}{lllrrrrrrrrrrrc}
\hline  \hline
  Sample&$\sigma$ & $P$& $x_P$&$D_{\mathrm{max}}$ &$x_{\mathrm{max}}$&
                                                                       $N_{\mathrm{max}}$&
                $L_g$ & $L_f$ &$D_t$  &$x_t$ &$N_{scl}$ & $L_g$ & $L_f$ & $F_{f}$\\  
\hline  
(1)&(2)&(3)&(4)&(5) &(6)&(7)&(8)&(9)&(10)&(11)&(12) &(13)&(14)&(15)\\ 
\hline  \\
  LCDM.0&0.6458 & 2.00 &1.55& 2.70 &2.63&8321  &  316 &178&
                                                       4.20&4.96 & 6044&  113  &142& 0.00788\\
  LCDM.1&0.3683 & 1.60 &1.63& 2.10 &2.99&8472 &  317& 178 & 
                      2.60&4.34  &  6524 &   113  &  150 & 0.00760\\
  LCDM.3&0.1852  & 1.30 &1.62& 1.50 &2.70&8535 &  348  & 190 &
                                                              1.70&3.78  &  6607 &   118 & 152 & 0.00930\\
  LCDM.10&0.0667& 1.10 &1.50& 1.16&2.40&8643&   332  & 174 &
                                                            1.20&3.00  &  7833 &   137 & 149 & 0.01469\\
   LCDM.30&0.0237& 1.04 &1.52& 1.06&2.36&8926&   312  & 167 &
                                                            1.06&2.61
                                             &  8582 &   143 & 142 &
                                                                     0.02137 \\
\\
  HCDM.0  & 0.8475 &2.10 &  1.30& 3.20 &  2.60 & 8158 & 288 & 173 &5.80 &  5.66  &  5109  &
                                                                       106  &  148 & 0.00591\\
  HCDM.1  & 0.4527  &1.70 & 1.55& 2.30 &  2.87 & 8342 &  314 & 177 &3.20   & 4.86&5855   &
                                                                         119   & 149 & 0.00727\\
  HCDM.3  & 0.2035  &1.30 & 1.47& 1.60&   2.95 & 8513 &  312 & 169&1.80   & 3.93&6580   &
                                                                        118   & 152 & 0.00895\\
  HCDM.10 & 0.0689 &1.10& 1.45& 1.18 &  2.61 & 8686 &  299 & 168&1.20   & 2.90&8148   &
                                                                      140   & 144 & 0.01750\\
  HCDM.30 & 0.0239 &1.04& 1.59& 1.06 &  2.42 & 8971 &  301 & 167&1.06   & 2.59&8664   &
                                                                      152   & 143 & 0.02251\\
\\
  OCDM.0 & 0.6548& 2.00 &1.53&   2.70&   2.60 & 8432 & 316 & 177&4.20  &4.89&  6112  &
                                                                   115  &  141 & 0.00818\\
  OCDM.1 & 0.4145& 1.70 & 1.69&   2.30&   3.14 & 8513 & 315 & 179&2.90  & 4.58& 6304  &
                                                                   124  &  147 & 0.00845\\
  OCDM.3 & 0.2489 &1.40 & 1.61&   1.70&   2.81 & 8581 & 319 & 181&1.90  &3.62&  7463   &
                                                                    137   & 144 & 0.01314\\
  OCDM.10 & 0.1039 & 1.16 & 1.54&   1.28&   2.69 &8752 & 298 & 168&1.32  &3.08&  7983   &
                                                                     137  &  145 & 0.01568\\
  OCDM.30 & 0.0404 & 1.06 & 1.54&   1.10&   2.43 &8936 & 301 & 167&1.11  &2.62&  8525   &
                                                                     143  &  142 & 0.02118\\
\\
  SCDM.0 & 0.5124 &1.60 & 1.17& 2.35 & 2.63&14808  & 194 & 125 & 2,65 &3.22&  14259  &
                                                                       130  &  106 & 0.02376\\
  SCDM.1 & 0.2600 &1.35 & 1.35&  1.70 &   2.69 & 15202 &  148  & 109  &1.85  &3.27& 14052   &
                                                                         106   & 106 & 0.01849\\
  SCDM.3  & 0.1314 &1.18 & 1.37&  1.34 &   2.59 &15160 &  144  & 111 &1.38  &  2.89  & 14406   &
                                                                        110   & 105 & 0.02106\\
  SCDM.10 & 0.0478 &1.07 & 1.46&  1.12 &   2.41 &15029 &  146  & 112 &1.13  & 2.72&14360   &
                                                                        142   & 105 & 0.02019\\
 SCDM.30 & 0.0170 &1.03 & 1.53&  1.04 &   2.35 &15192 &  146  & 107 &1.04  & 2.59&14756   &
                                                                        145   & 105 & 0.02164\\
\\
 SDSS &    &2.5 && 3.5 &&1129 & 249 & 147&5.40 &&    844 &   118 &   134 & 0.00981
\label{Tab2}                         
\end{tabular} 
\tablefoot{
Table columns are
\noindent (1): the sample name, where the last number shows the redshift;
\noindent (2):   $\sigma$ the standard deviation of the density field;
\noindent (3): $P$ the percolation  density threshold in mean density units; 
\noindent (4):  $x_P=(P-1)/\sigma$ the reduced percolation  density threshold; 
\noindent (5): $D_{\mathrm{max}}$ the density threshold at the maxima of supercluster numbers;
\noindent (6): $x_{\mathrm{max}}=(D_{\mathrm{max}}-1)/\sigma$ the reduced density threshold at the maxima of supercluster numbers;
\noindent (7): $N_{\mathrm{max}}$ the maximum number of superclusters;
\noindent (8):  $L_g$ the geometrical diameter (length) of the largest supercluster
in \Mpc\ at $D_{\mathrm{max}}$;
\noindent (9):  $L_f$ the fitness diameter (length) of the largest supercluster
in \Mpc\ at $D_{\mathrm{max}}$;
\noindent (10): $D_t$ the density threshold for finding
superclusters in mean density units; 
\noindent (11): $x_t = (D_t -1)/\sigma$ the  reduced density threshold for finding
superclusters; 
\noindent (12): $N_{scl}$ the number of superclusters at $D_t$;
\noindent (13):  $L_g$ the geometrical diameter (length) of the largest supercluster
in \Mpc\ at $D_t$;
\noindent (14):  $L_f$ the fitness diameter (length) of the largest supercluster
in \Mpc\ at $D_t$;
and \noindent (15): $F_f$ the  total filling factor of over-density regions at
$D_t$.
}
\end{table*} 
}

\subsection{Supercluster fitness diameters}

Following \citet{Einasto:2019fk}, we defined the fitness volume of the
supercluster, $V_f$, to be proportional to its geometrical volume,
$V_{g}$, and divided by the total filling factor:
\begin{equation}
  V_f =V_{g}/F_f.
  \label{fitness0}
\end{equation}
Using the definition of the total filling factor  of all over-density
regions at this threshold density, $F_f = V_C/V_0$, we obtain
\begin{equation}
  V_f = V_g/V_C \times V_0.
  \label{fitness}
\end{equation}
The fitness volume measures the ratio of the supercluster volume to
the summed volume of all superclusters (all filled over-density
regions) at the particular threshold density, multiplied by the whole
volume of the sample.  Fitness diameters (lengths) of superclusters
are calculated from their fitness volumes,
\begin{equation}
L_f = V_f^{1/3} = (V_g/V_C)^{1/3} \times L_0. 
\end{equation}
We used the fitness diameters of the largest superclusters, $L_f(D_t)$, as a
percolation function in addition to other percolation functions such
as geometrical diameters, $L_g(D_t)$, total filling factors,
$F_f(D_t)$, 
and numbers of clusters, $N(D_t)$.  The fitness diameters of the largest
superclusters are functions of the threshold density $D_t$ and have a
minimum at a medium threshold density.  This minimum shows that the
largest supercluster has the smallest volume fraction, $V_g/V_C$. The
minimum was used to find the threshold density for the supercluster
selection.  We considered the fitness volume of a supercluster as the
volume of its basin of dynamical attraction or cocoon
\citep{Einasto:2019fk}.  The sum of the fitness volumes of 
the supercluster  cocoons is equal to the volume of the sample:
$\sum V_f = \sum V_g/V_C \times V_0 = V_0$.

\section{Evolution of superclusters as described by percolation
  functions}

We discuss is this section the evolution of superclusters as described
by percolation functions. Next we analyse the evolution of
distributions of the supercluster diameters and luminosities, and the errors
of the percolation parameters.

\subsection{Evolution of the percolation functions of  the model samples}

We used the percolation functions to characterise the geometrical properties of
the cosmic web and to select superclusters.  Fig.~\ref{fig:evol} shows
geometrical length functions, $L_g$, the fitness diameter functions,
$L_f$, and the numbers of clusters, $N$. The upper panels show these functions
for the LCDM model and in the following panels for the HCDM, OCDM and SCDM
models for redshifts $z=0,~1,~3,~10,\text{and}~30$.  An important parameter is
the standard deviation (rms variance) of the density contrast, $\sigma$,
which was calculated using Eq.~(\ref{disp}) for all our models. The results are
given in Table~\ref{Tab2}.  In Fig.~\ref{fig:evol} we use the reduced
threshold density $x= (D_t - 1)/\sigma$ as the argument of the percolation
functions.

An essential indicator of the cluster evolution is their number.
Fig.~\ref{fig:evol} shows the number of clusters as a function of the
threshold density. At very low threshold densities, the whole
over-density region contains one percolating cluster because the
density field peaks are connected by filaments to a connected region.
For this reason, one percolating cluster exists at a low threshold
density, $x \le 1.5$ that extends over the whole volume of the
computational box.  The percolation threshold density, $P = D_t$, is
defined as follows: for $D_t \le P,$ one and only one percolating
cluster exists, and for $D_t > P,$ no percolating clusters exist
\citep{Stauffer:1979aa}. We denote the percolation threshold in
reduced threshold density units as $x_P$.  At these low threshold
densities, the geometrical diameter of the cluster is equal to the
diameter of the box, $L_g = \sqrt{3}~L_0$, and its fitness diameter is
equal to the side length of the box, $L_f=L_0$.

With increasing threshold density, some filaments became fainter than
the threshold density, and the connected region split into smaller
units, supercluster candidates and their complexes.  This led to a
rapid increase in the number of clusters with increasing threshold
density at $x > -0.5$.  At $x \approx 2.6$ (for the LCDM.0 model), the
number of clusters reached a maximum, $N_{\mathrm{max}} \approx 8300$.
The threshold density at the maximum number of clusters,
$D_{\mathrm{max}}$ and $x_{\mathrm{max}}$,  the respective numbers of
clusters, $N_{\mathrm{max}}$, and the geometrical and fitness diameters, $L_g$
and $L_f$, are given in Table~\ref{Tab2}. At this threshold density,
most clusters are still complexes of large over-density regions,
connected by filaments to form systems with diameters of
$L_g \approx 300$~\Mpc\ and $L_f \approx 200$~\Mpc.

\begin{figure*}[ht]
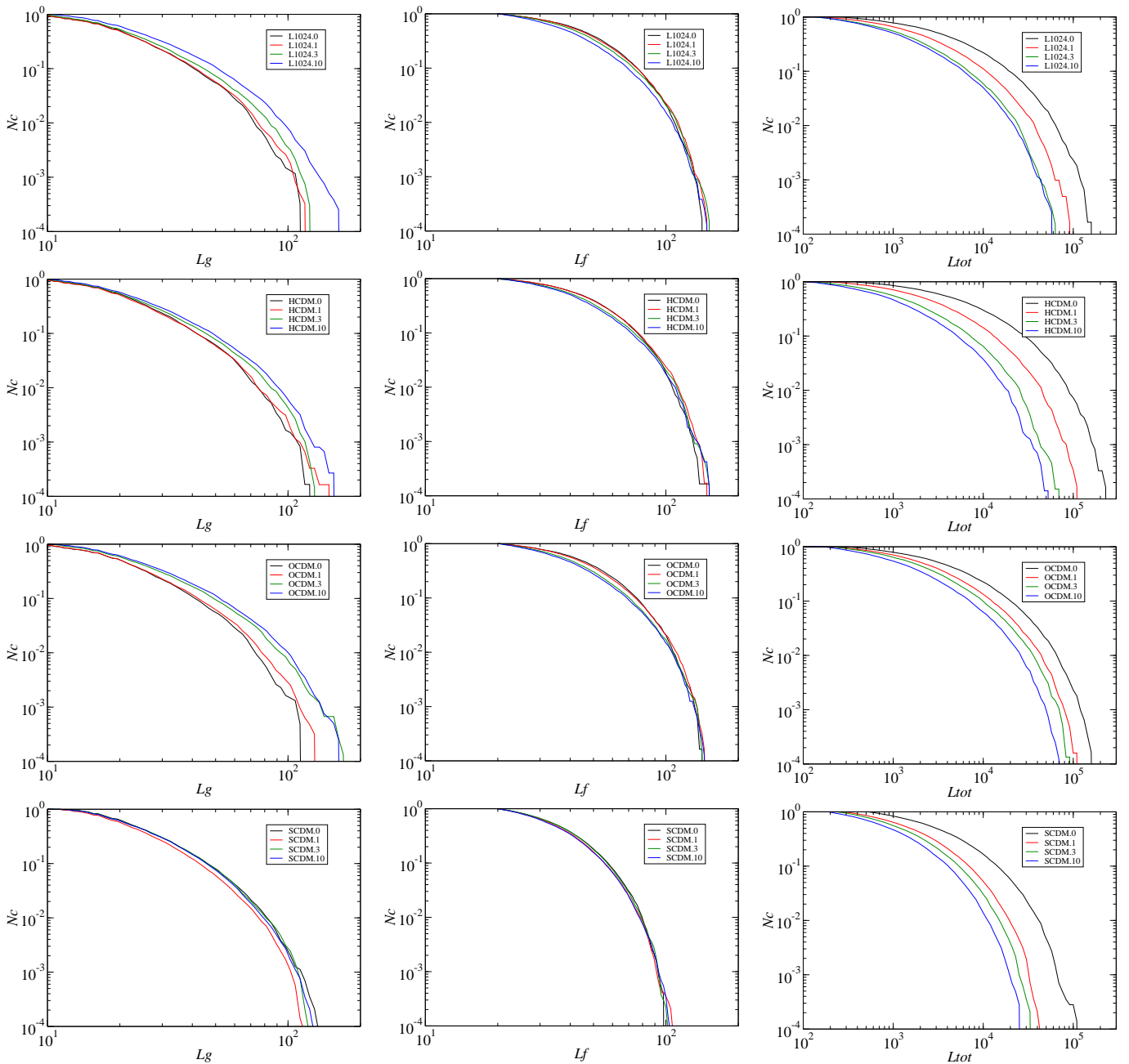
 
\centering 
\hspace{2mm} 
\resizebox{0.31\textwidth}{!}{\includegraphics*{LCDM_geomdiam-distr.eps}}
\hspace{2mm} 
\resizebox{0.31\textwidth}{!}{\includegraphics*{LCDM_dyndiam-distr.eps}}
\hspace{2mm}  
\resizebox{0.31\textwidth}{!}{\includegraphics*{LCDM_lum-distr.eps}}\\
\hspace{2mm} 
\resizebox{0.31\textwidth}{!}{\includegraphics*{HCDM_geomdiam-distr.eps}}
\hspace{2mm} 
\resizebox{0.31\textwidth}{!}{\includegraphics*{HCDM_diam-distr.eps}}
\hspace{2mm}  
\resizebox{0.31\textwidth}{!}{\includegraphics*{HCDM_lum-distr.eps}}\\
 \hspace{2mm} 
\resizebox{0.31\textwidth}{!}{\includegraphics*{OCDM_geomdiam-distr.eps}}
\hspace{2mm} 
\resizebox{0.31\textwidth}{!}{\includegraphics*{OCDM_diam-distr.eps}}
\hspace{2mm}  
\resizebox{0.31\textwidth}{!}{\includegraphics*{OCDM_lum-distr.eps}}\\
\hspace{2mm} 
\resizebox{0.31\textwidth}{!}{\includegraphics*{SCDM_geomdiam-distr.eps}}
\hspace{2mm} 
\resizebox{0.31\textwidth}{!}{\includegraphics*{SCDM_diam-distr.eps}}
\hspace{2mm}  
\resizebox{0.31\textwidth}{!}{\includegraphics*{SCDM_lum-distr.eps}}
\caption{Cumulative distribution of supercluster geometrical
  diameters, $L_g$ ({\em left} panels), fitness diameters, $L_f$ ({\em
    middle} panels), and total masses, $\cal{L}$ ({\em right}
  panels). The distributions are normalised to the total
  supercluster numbers. The panels from the {\em top} to the {\em bottom} show the
  LCDM, HCDM, OCDM, and SCDM models. }
\label{fig:Fig3} 
\end{figure*} 

When we increase $x$ more,  the number of clusters starts to
decrease because the smallest clusters have maximum densities that are lower than
the threshold density, and they disappear from the sample.  At
$x \approx 4,$ the geometrical and fitness diameters become equal,
$L_g \approx D_d \approx 160$~\Mpc. With a further increase of the
density threshold geometrical diameters decrease, but the fitness
diameters have a minimum and thereafter start to increase.  As shown
in Fig. \ref{fig:evol} and Table~\ref{Tab2}, the minimum fitness
diameters are almost identical (in co-moving coordinates) at all
epochs, $L_f \approx 140$~\Mpc\ for the LCDM model. The geometrical
diameter at this threshold density is $L_g \approx 115$~\Mpc. Both
diameters are close to the conventional values of supercluster
diameters.  We used threshold densities at global minima of 
the fitness diameter functions to select supercluster ensembles.
The parameters of the model supercluster samples at these threshold densities
are given in Table~\ref{Tab2}. They are $D_t$, $x_t$, $N_{scl}$, $L_g$ , and
$L_f$.  The table also lists the total filling factor of
over-density regions, $F_{f}$, at the threshold density $D_t$.  This
filling factor was used to find fitness volumes and diameters of
superclusters, see Eq.~(\ref{fitness}).  Data are given for all our
model samples and evolutionary epochs.

The decrease of the number of clusters with increasing threshold
continues  until only the central
cluster regions have densities that are higher than the threshold
density. For the earliest epoch $z=30,$ the decrease in diameters with
increasing threshold density is fastest (the diameters are expressed in
co-moving coordinates).

\begin{figure*}[ht]
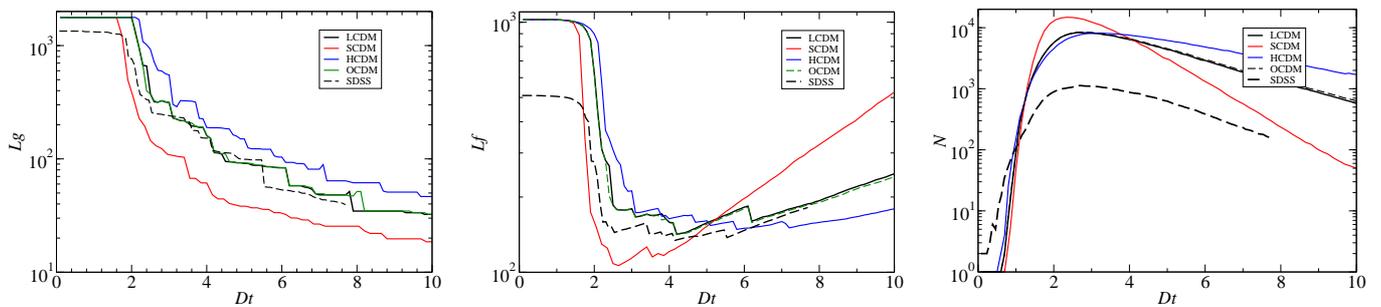
 
\centering 
\hspace{2mm}
\resizebox{0.31\textwidth}{!}{\includegraphics*{LCDM_Lgeom_Dtrial.eps}}
\hspace{2mm} 
\resizebox{0.31\textwidth}{!}{\includegraphics*{LCDM_Ldyn_Dtrial.eps}}
\hspace{2mm} 
\resizebox{0.31\textwidth}{!}{\includegraphics*{LCDM_N_Dtrial.eps}}\\
\caption{Comparison of the percolation functions of models with SDSS
  samples. The model functions are plotted with coloured bold lines.
  The functions for SDSS samples are plotted with dashed black lines for
  the density correction factor   1.30. The {\em left panel} shows geometrical length
  functions, the {\em middle panel} the fitness length functions, and the {\em right panel}
  the number functions.}
\label{fig:Fig4} 
\end{figure*}

Fig.~\ref{fig:evol} and Table~\ref{Tab2} show that the maximum
numbers of clusters are very similar at all evolutionary stages of the
cosmic web.  This similarity, as well as the similarity of the minimum
fitness diameters at different epochs, is an important property of the
evolution of the cosmic web.

\subsection{Diameter and mass distributions}

Fig.~\ref{fig:Fig3} shows the cumulative distributions of the
geometrical and fitness diameters and supercluster masses.  The data
are given for all models and simulation epochs up to
$z=10$. Fig.~\ref{fig:Fig3} shows that the geometrical diameters at
early epochs are larger than at the present epoch (in co-moving
coordinates) by a factor of approximately 2.  This effect is seen in
the LCDM, OCDM, and HCDM models, but it is almost absent in the SCDM
model. This means that superclusters shrink during the evolution in
co-moving coordinates.  The fitness diameters have a different
behaviour: the distribution of the fitness diameters is almost the
same in co-moving coordinates at all epochs.

The right panels of Fig. ~\ref{fig:Fig3} show that the supercluster
masses increase during the evolution by a factor of approximately 3.
This result agrees well with all simulations of 
the cosmic web growth. The skeleton of the web with
supercluster embryos already forms at an early epoch.  Superclusters grow by the
infall of matter from low-density regions towards early forming knots
and filaments, forming early superclusters.  The mass growth is
largest in the HCDM model.

\subsection{Errors on the percolation parameters}

As shown by \citet{Einasto:2019fk}, percolation parameters depend on
the smoothing length that is used to calculate the density field. { The
changing of the smoothing length of the density field allows us to
select systems of galaxies of various character, see below. For this
reason, the errors of the percolation parameters characterise the
accuracy of parameters in relation to galaxy systems for a given
smoothing length.}  Some percolation functions are very smooth, and the
possible parameter error is given by the step size of threshold
density, which is $\Delta D_t=0.1$ in the present study in most
cases. This determines the parameter accuracy: the percolation
threshold density $P$, the density threshold at maxima of supercluster
numbers $D_{\mathrm{max}}$, and the density threshold for finding
superclusters $D_t$.  For early epochs the density field was scanned
with a smaller step, and the possible errors on these parameters are
lower. The errors on other parameters depend on the speed of the
parameter changes as functions of the density threshold.  The possible
error range can be estimated from the parameter spread for different
models and evolution epochs in Figs. \ref{fig:Fig6} and
\ref{fig:Fig7}.

\begin{figure*}[ht]
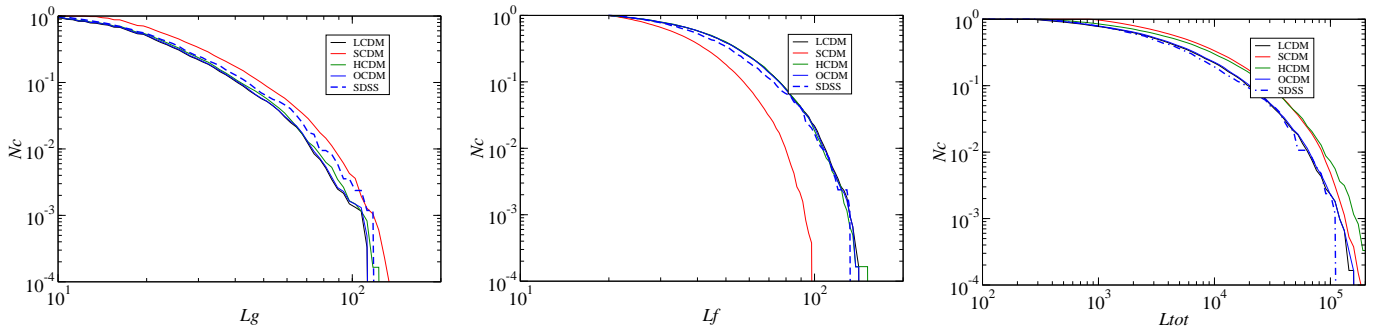
 
\centering 
\hspace{2mm} 
\resizebox{0.31\textwidth}{!}{\includegraphics*{LCDM_SDSSgeomdiam-distr.eps}}
\hspace{2mm} 
\resizebox{0.31\textwidth}{!}{\includegraphics*{LCDM_SDSSdyndiam-distr.eps}}
\hspace{2mm}  
\resizebox{0.31\textwidth}{!}{\includegraphics*{LCDM_SDSSlum-distr1.45.eps}}
\caption{Comparison of the cumulative diameter and mass (luminosities)
  distribution of models at the present epoch and the SDSS
  samples. The {\em left} panel shows the cumulative distributions of
  supercluster geometrical diameters $L_g$, the {\em middle} panel
  shows the distributions of the fitness diameters $L_f$, and the {\em
    right} panel shows the total mass (luminosity) distributions
  $\cal{L}$, given in units of the mass (luminosity) of one cell. The
  SDSS distributions are given for a threshold density $D_t =5.4$, and
  the distribution of SDSS total luminosities is calculated for a
  correction factor $b=1.45$. }
\label{fig:Fig5} 
\end{figure*}

\section{Discussion}

We start with the comparison of model percolation functions with
percolation functions of  observed SDSS samples.  Next we discuss the
evolution of the ensemble of superclusters in models with different
cosmological parameters.  Then we compare the evolution of LCDM and
SCDM models, and evolution of LCDM, OCSM and HCDM models. Thereafter
we discuss the influence of smoothing length to properties of selected
systems, and the concept of cocoons of the cosmic web.  Finally we
compare our analysis with results of the study of the velocity field
to detect supercluster cocoons.

\subsection{Comparison with SDSS samples}

In Fig.~\ref{fig:Fig4} we compare percolation functions of observed
SDSS samples with the percolation functions of the models at the present
epoch.  We note that there are approximately eight times more model
superclusters than SDSS superclusters.  This difference 
is due to the larger size of our model samples, 1024~\Mpc, which is  
about twice 
the effective size of the SDSS main galaxy sample, 509~\Mpc.  To bring
percolation functions of SDSS samples to the same scale as that model
functions, the threshold densities of the SDSS samples must be shifted.  A
similar shift in the density threshold of the SDSS samples was made by
\citet{Einasto:2019fk}. The densities are expressed in mean density units.
In the model samples, the mean density includes DM  in
low-density regions in addition to the
clustered matter with simulated galaxies.  Low-density regions include no simulated
galaxies, or the galaxies are fainter than the magnitude limit of the
observational SDSS survey.  For this reason, unclustered and low-density
DM is not included in calculations of the
mean density of the observed SDSS sample.  This means that in the calculation of
densities in mean density units, the densities are divided into a smaller
number, which increases the density values of the SDSS
samples. \citet{Einasto:2019fk} estimated this correction factor by a
trial-and-error procedure and calculated corrected threshold
densities by dividing the threshold densities of the SDSS samples by the
factor $b=D_t/(D_t)_c $.  We applied the same factor $b=1.30$ and
used it to calculate SDSS percolation functions for comparison.  The
corrected SDSS supercluster diameter, filling factor, and number
functions agree well with the LCDM model functions.

Fig. ~\ref{fig:Fig5} shows the cumulative distributions of
the diameters and masses of the model samples at the present epoch and the
respective distributions for the SDSS samples.  To calculate the
luminosity distribution of the SDSS superclusters, we divided the
supercluster luminosities into the  normalising factor $b=1.45$,
following 
\citet{Einasto:2019fk}.  The right panel in
Fig.~\ref{fig:Fig5} shows that this correction brings the total luminosity
distributions of the SDSS and LCDM samples to a very good agreement.
The diameter and luminosity distributions are shown in
Figs.~\ref{fig:Fig3} and \ref{fig:Fig5}.  The
distributions of the different models clearly have an approximately similar
character. 

\subsection{Evolution of the percolation parameters of supercluster
  ensembles} 

We showed in Fig.~\ref{fig:evol} that the basic characteristics
of the evolution of the percolation functions in different cosmologies are
rather similar.  At the earliest epoch $z=30,$ the percolation functions,
expressed as functions of the reduced threshold densities $x$, are almost
identical for models with different cosmological parameters.  The maxima
of the number functions and the minima of the fitness length functions are located
in all models at the earliest epoch $z=30$ at a reduced threshold density
$x_{\mathrm{max}} = 2.4$.  At the present epoch $z=0,$ the maximum
shifts to $x_{\mathrm{max}} = 2.6$ for all models.  The shape of
the fitness length and the number functions at early epoch is approximately
symmetrical around $x=x_{\mathrm{max}}$ when expressed as a function
of the reduced threshold density $x$.  It is surprising that in spite
of different values of $\sigma$ at the earliest epoch $z=30,$
the percolation functions of all models at early epoch are so similar. At
later epochs, this symmetry of the percolation functions is better
preserved in the LCDM, HCDM, and OCDM models. The evolution of the 
SCDM model is different: at later epochs, the percolation functions of the SCDM
model are shifted far more strongly to higher threshold densities than in
other models, see Figs.~\ref{fig:Fig4} and \ref{fig:Fig5}.

\begin{figure*}[ht]
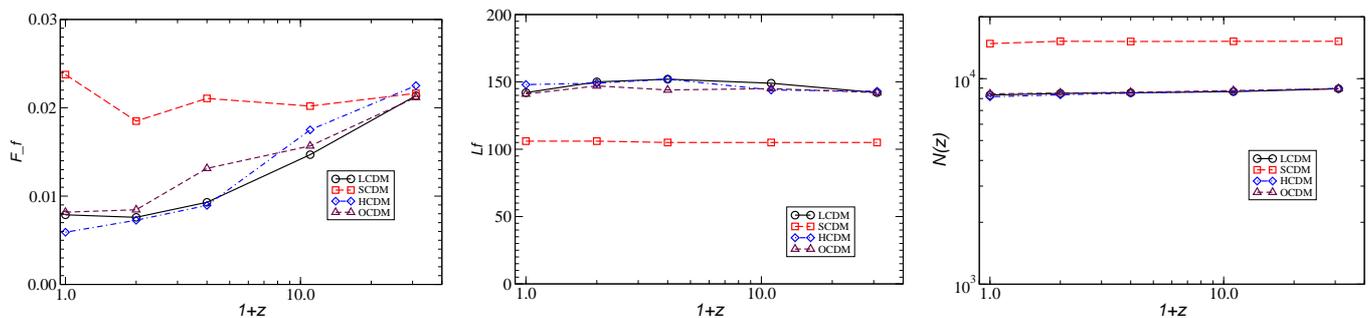
 
\centering 
\hspace{2mm}
\resizebox{0.31\textwidth}{!}{\includegraphics*{evol_Ff-z.eps}}
\hspace{2mm} 
\resizebox{0.31\textwidth}{!}{\includegraphics*{evol_Lf-z.eps}}
\hspace{2mm} 
\resizebox{0.31\textwidth}{!}{\includegraphics*{evol_Nmax-z.eps}}
\hspace{2mm} 
\caption{ { Evolution of percolation parameters of supercluster
  ensembles. } {\em Left panel:} Change in filling factor $F_f$ in the
  models with cosmic epoch $z$. {\em Middle panel:} Evolution of the
  minimum fitness lengths with epoch $L_f(z)$.  {\em Right} panel:
  Evolution of the maximum numbers of clusters with epoch $N(z)$.  }
\label{fig:Fig6} 
\end{figure*}

Now we discuss the evolution of the percolation parameters of the ensemble
of superclusters in more detail.  In Fig.~\ref{fig:Fig6} we show the
change in three percolation parameters of supercluster ensembles with
cosmic epoch $z$: the filling factor $F_f$, 
the minimal fitness lengths $L_f$, and the maximum number of
superclusters $N_{\mathrm{max}}$.  
The left panel of Fig.~\ref{fig:Fig6} shows the change in the filling factor $F_f$ of
the models during the evolution. This filling factor was used to calculate
the fitness volumes of superclusters $V_f$ using Eq.~(\ref{fitness0}).
At the earliest epoch $z=30,$ the filling factor of the superclusters in all
models was $F_f \approx 0.02$.  During the evolution, the filling
factor decreased in the LCDM, HCDM, and OCDM models to $F_f \approx 0.007$,
but it remained almost the same  $F_f \approx 0.02$ for the SCDM model.

\begin{figure*}[ht]
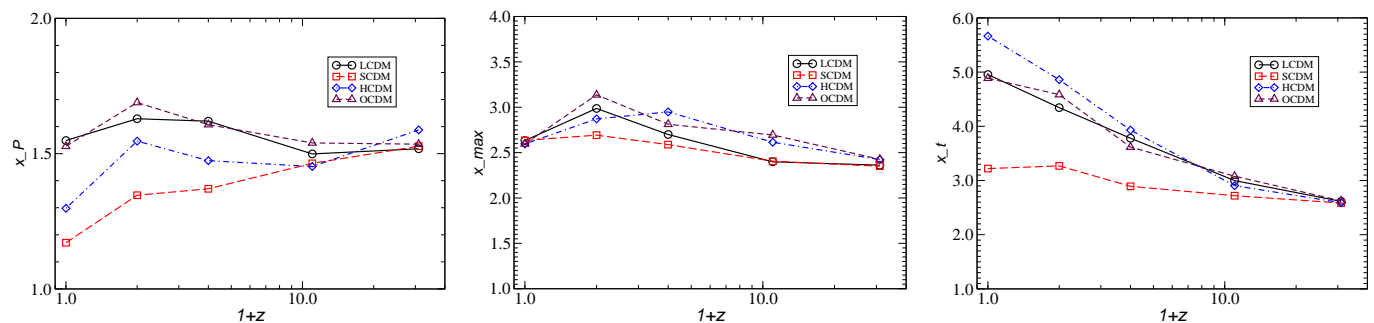
 
\centering 
\hspace{2mm}
\resizebox{0.31\textwidth}{!}{\includegraphics*{evol_xP-z.eps}}
\hspace{2mm} 
\resizebox{0.31\textwidth}{!}{\includegraphics*{evol_xmax-z.eps}}
\hspace{2mm} 
\resizebox{0.31\textwidth}{!}{\includegraphics*{evol_xt-z.eps}}
\hspace{2mm} 
\caption{ { Evolution of reduced percolation parameters of supercluster
  ensembles.}  {\em Left panel:} Evolution of the reduced percolation
  threshold with cosmic epoch. {\em Middle panel:} Change in evolution
  of the reduced density threshold at the maxima of the numbers of
  superclusters. {\em Right panel:} Evolution of the reduced density
  threshold for finding superclusters.  }
\label{fig:Fig7} 
\end{figure*}

In the middle and right panels of Fig.~\ref{fig:Fig6} we show the minimum
fitness diameters and maximum numbers of superclusters as functions of
the cosmic epoch.  The LCDM, HCDM, and OCDM models have almost
identical evolutions of the sizes and supercluster numbers. The maximum
length of the fitness diameter of the SCDM model is about 100~\Mpc, which is a
factor of about 1.5 times smaller than in other models,
$L_f \approx 140$~\Mpc.  The maximum number of SCDM superclusters is
almost twice as high as the number in other models.  The difference of the SCDM
model from other models in the maximum number of superclusters and in
the minimum fitness length of SCDM superclusters is an essential
finding of the present paper.  We conclude that the structure of the
ensemble of SCDM superclusters differs considerably from the structure
of the ensemble of superclusters in other models.

Fig. ~\ref{fig:Fig7} shows the evolution of three parameters of
the ensembles of models: the reduced percolation threshold $x_P$, the
reduced density threshold at the maxima of the numbers of superclusters
$x_{\mathrm{max}}$, and the reduced density threshold for finding
superclusters $x_t$.  The reduced percolation lengths of the LCDM models of
different box sizes and smoothing scales have mean values
$x_P =1.5 \pm 0.1$ for all simulation epochs \citep{Einasto:2019fk}.
Our study shows that models with different cosmology also have a
reduced percolation length $x_P \approx 1.5$ for the LCDM, HCDM, and OCDM
models for all cosmic epochs, which confirms early results by
\citet{Colombi:2000fj}.  The SCDM model has the value $x_P =1.5$ only
for the earliest epoch $z=30$. At later epochs, the reduced percolation
threshold is lower, see the left panel of Fig.~\ref{fig:Fig7}.

The reduced density threshold at the maxima of the number of superclusters
is $x_{\mathrm{max}} = 2.4$ for all models at the earliest epoch
$z=30$.  During the evolution, the reduced density threshold at the maxima
of the number of superclusters increases to $x_{\mathrm{max}} = 2.6$
for all models.  The reduced threshold density at the minimum of the
fitness length (optimal to select superclusters) is at the earliest
epoch $x_t = 2.6$ for all models.  It increases to a value $x_t = 5.0$
for the LCDM and OCDM models, and to $x_t = 5.7$ for the HCDM model, but
it remains almost the same, $x_t = 3.2$, for the SCDM model. 

\begin{figure*}[ht] 
\centering 
\hspace{2mm}
\resizebox{0.45\textwidth}{!}{\includegraphics*{evol_ds-z.eps}}
\hspace{2mm}
\resizebox{0.48\textwidth}{!}{\includegraphics*{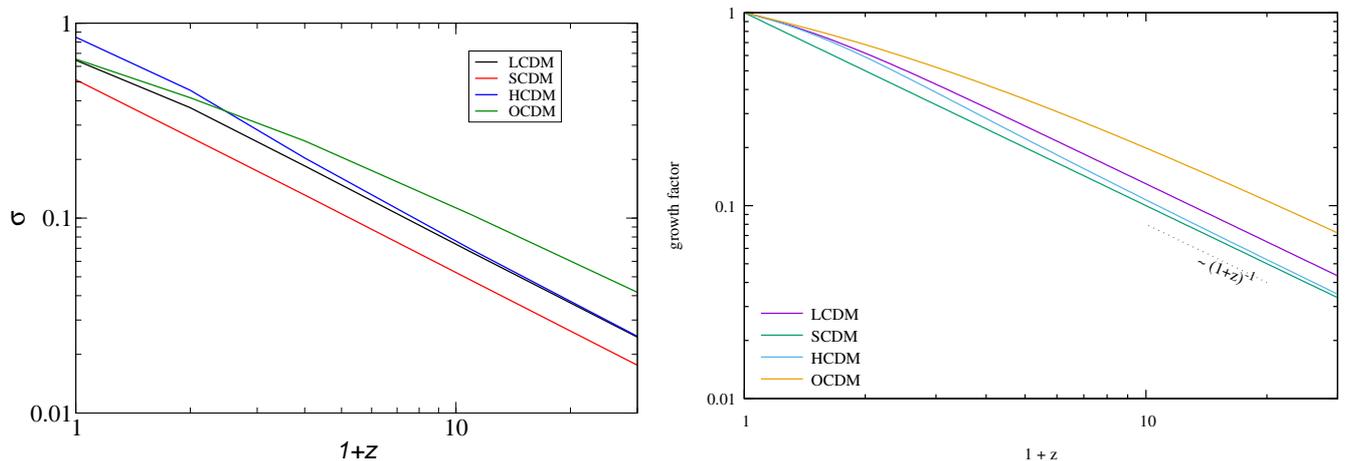}}
\caption{ { Evolution of the standard deviation (dispersion)  of density fluctuations.}
  {\em Left panel:}  Change in dispersion of the density
  fluctuations $\sigma$ with cosmic epoch $z$ for our models. {\em
    Right panel:}  Evolution of the dispersion of the density
  fluctuations $\sigma$ with cosmic epoch $z$  according to the linear
  evolution model, normalised to the present epoch. 
   } 
\label{fig:Fig8} 
\end{figure*}

\citet{Einasto:2019fk}
investigated the evolution of the standard deviation in $\Lambda$CDM
(LCDM) models of various box lengths and different smoothing scales.
The authors showed that the shape of the relation between the density
contrast $\sigma$ and redshift $1+z$ is approximately linear when
expressed in log-log format.  The slope of the relation is the
same for the LCDM models of different box lengths and smoothing scales,
and the amplitude depends on the smoothing scale.  In the present
paper we used identical box sizes and smoothing scales, but varied
the cosmological model parameters.
The variance of the standard deviation $\sigma$ as a function of the
cosmic epoch $z$ is shown in the left panel of
Fig.~\ref{fig:Fig8}. The relation
between $\sigma$ and $1+z$ is almost linear when expressed in log-log
format, which agrees well with linear perturbation theory; see the
right panel 
of Fig.~\ref{fig:Fig8}.  The HCDM model has a more rapid increase of
the standard deviation $\sigma$ with time (decreasing $z$), and the OCDM
model has the slowest increase of the standard deviation with time.

\subsection{Comparison of the evolution of the LCDM and SCDM models}

\begin{figure*}[ht] 
\centering 
\resizebox{0.95\textwidth}{!}{\includegraphics*{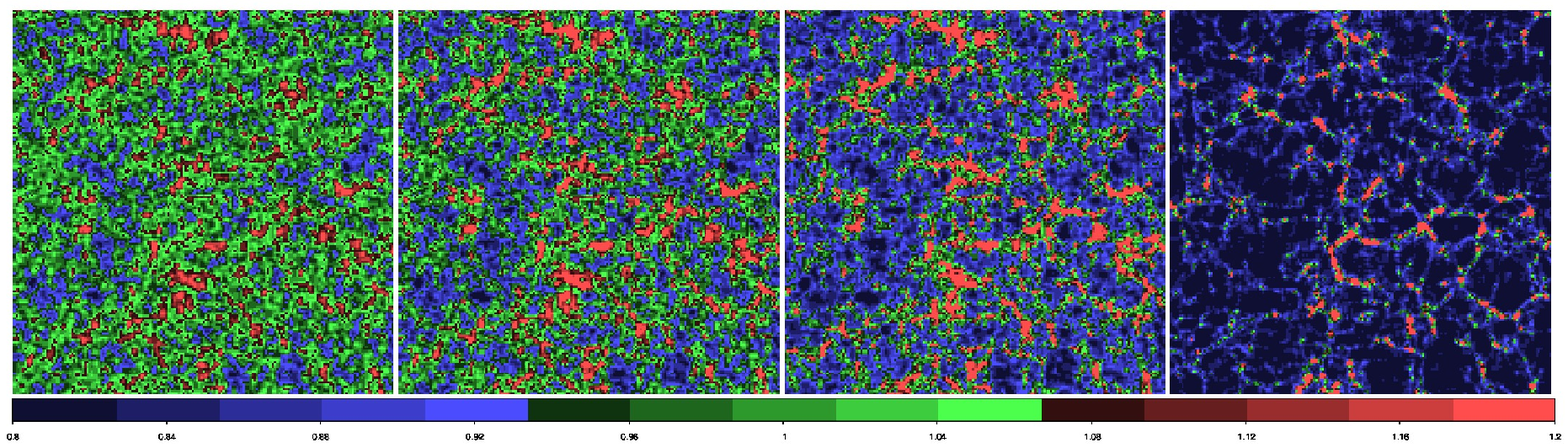}}

\resizebox{0.95\textwidth}{!}{\includegraphics*{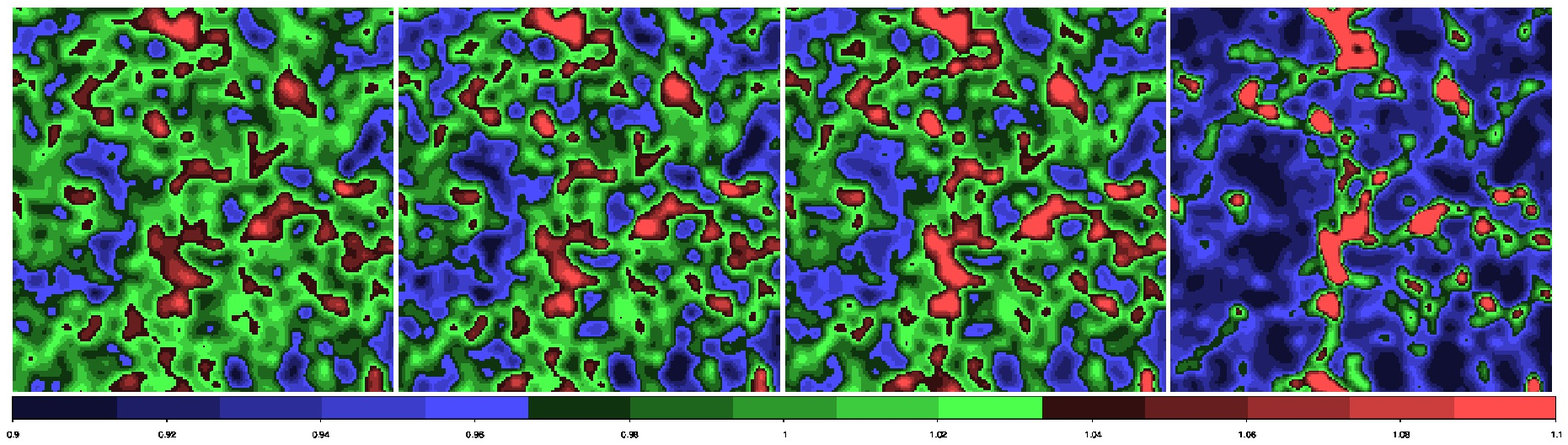}}

\resizebox{0.95\textwidth}{!}{\includegraphics*{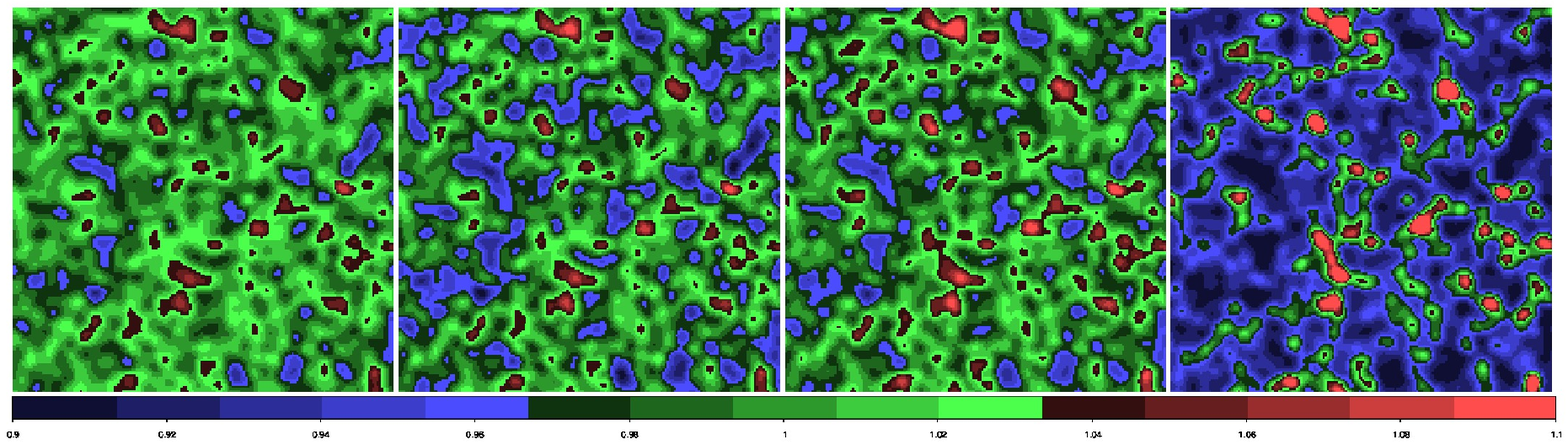}}
\caption{ { Evolution of the density fields of LCDM and SCDM models.}
  {\em Upper} panels: Density fields of the LCDM model found with a
  smoothing kernel of radius 2~\Mpc. {\em Middle} and {\em lower}
  panels: Density fields of the LCDM and SCDM models found with a
  smoothing kernel with a radius 8~\Mpc.  The panels from left to
  right correspond to epochs $z=30,~10,~3,\text{and}~0$.  Cross
  sections are shown in a 2~\Mpc\ thick layer of size
  $341 \times 341$~\Mpc. Densities are expressed on a linear
  scale. Colour scales from left to right in the upper panel are
  $0.8 - 1.2,~0.6 - 1.45,~0.3 - 2,~\text{and }0.1 -5.0$, and in the
  middle and lower panels, they are
  $0.9 - 1.1,~0.8 - 1.25,~ 0.4 - 1.65,\text{and } 0.2 - 3.0$.}
\label{fig:Mdenfield} 
\end{figure*}

Our  analysis has shown that the properties of the SCDM model
deviate strongly from the properties of other models. The main reason for
this difference lies in the power spectrum of the SCDM model: it
has much more power on small scales and less power on large scales.
The evolution of the ensemble of superclusters in our SCDM and LCDM models 
can be followed in Fig.~\ref{fig:Mdenfield}. The upper and middle panels of this
figure show cross sections of the density fields of the LCDM model and the lower
panels show the SCDM model.  The panels from left to right correspond to
density fields at epochs $z=30,~10,~3,\text{anda}~0$, calculated  with co-moving smoothing
kernel with a radius 8~\Mpc.    The models were calculated with identical
phases of the initial density fluctuations.  For this reason, the
small-scale features of the density fields of the different models are rather
similar. However, large-scale features on supercluster scales are
different.  In the SCDM model structures are smaller in size. To see
better details we plot in Fig.~\ref{fig:Mdenfield} only central $341
\times 341$~\Mpc\ sections of density fields. 

The left panels of Fig.~\ref{fig:Mdenfield} show the density fields of the models
for a very early epoch, corresponding to redshift $z=30$.  The
comparison of the panels of both models at identical epochs suggests that
the principal large-scale structural elements of the cosmic web were
present at all epochs considered here.  There are
differences on small scales, but the main large-scale elements of the web
are seen at similar locations at all epochs.  The basic visible changes
are the increase in density contrast: the density distributions
at epochs $z = 30$, $z = 10,$ and $z=3$ are very similar, only the
amplitude of density perturbations has increased.  In this redshift
range the evolution is mainly linear, only the density contrast has
increased.  In later epochs the non-linear evolution is dominant, as
seen by comparison of fields at $z=3$ and $z=0$.  In the SCDM model
the superclusters are smaller and their spatial density is higher than in
the LCDM model.

\subsection{Comparison of the evolution of the LCDM, OCDM, and HCDM models}

The differences in the evolution of the cosmic web are related  to the differences
in their initial power spectra.  Fig.~\ref{fig:spectr} shows that
the spectra of the LCDM, OCDM, and HCDM models are identical on medium and
small scales.  On the largest scales the open OCDM model has a larger power
than the conventional LCDM model and the hyper-DE HCDM model has
a lower power. Because differences in power spectra are very small, we
expect to observe also small differences in the geometrical properties
of the cosmic web, as represented in these models. The DE contribution to
the matter and energy density in early epochs was very small, which may
explain the low sensitivity of the cosmic web properties to the DE
density.  The HCDM model has the highest speed of the growth of the
amplitude of density fluctuations, $\sigma$, which may explains the 
largest growth of the supercluster masses during the evolution, as
shown in Fig.~\ref{fig:Fig3}.  For the same reason, the geometrical
diameters 
and numbers of HCDM superclusters at high threshold densities are
larger than for the LCDM and OCDM superclusters.

\subsection{Superclusters as great attractors in  the
  cosmic web} 

All massive bodies are gravitational attractors.   Galaxy-type and larger attractors are of interest in
cosmology.  It is well known 
that smoothing affects the character of the high-density regions that are found.
Smoothing with a kernel of length 1~\Mpc\ highlights ordinary galaxies
together with their satellite systems, similar to our Galaxy and M31.
These attractors can be called small in the cosmological context.  Within
their { spheres of dynamical influence}, galaxy-type attractors are
surrounded by dwarf satellites and intergalactic matter inside their
DM halos.  The central galaxies of these systems grow by infall of gas and
merging of dwarf galaxies; for a detailed overview, see
\citet{Wechsler:2018fj}. \citet{Einasto:1974aa} called these systems
hypergalaxies. The authors suggested that hypergalaxies are primary sites
of galaxy formation, and that galaxies do not form in isolation because
dwarf galaxies primarily exist as satellites of brighter (central)
galaxies.  The radius of satellite systems and of the DM halo that
surrounds the system, that is, hypergalaxies, is about 1~\Mpc.  For early
evidence, see \citet{Einasto:1974, Einasto:1974b}.

Smoothing with a kernel of length 4~\Mpc\ finds high-density regions
of the cosmic web that have an intermediate character between clusters and
traditional superclusters, such as central regions of superclusters
\citep{Einasto:2012aa, Einasto:2020cc}.  As shown by
\citet{Einasto:2019fk}, smoothing with a 4~\Mpc\ kernel finds four times
more isolated high-density systems than smoothing with an 8~\Mpc\ length
using the same LCDM model.  When a larger smoothing length of 16~\Mpc\ 
is used for this model, four times fewer supercluster-type
systems are found, and superclusters are larger than superclusters found
with the 8~\Mpc\ kernel.

Long experience in the study of superclusters on the basis of density
fields has shown that the optimal smoothing length for finding
superclusters is 8~\Mpc; see \citet{Einasto:2007tg}, 
\citet{Luparello:2011fr}, and \citet{Liivamagi:2012}.  Superclusters
are great attractors.  Supercluster centres lie at the centres of deep
potential wells.  They collect material from a much larger region than
clusters.  The slope of the potential field determines the speed of
particles at a given location. Thus superclusters at different levels of
the potential field have different strength  {  as} attractors.

\begin{figure*}[ht]
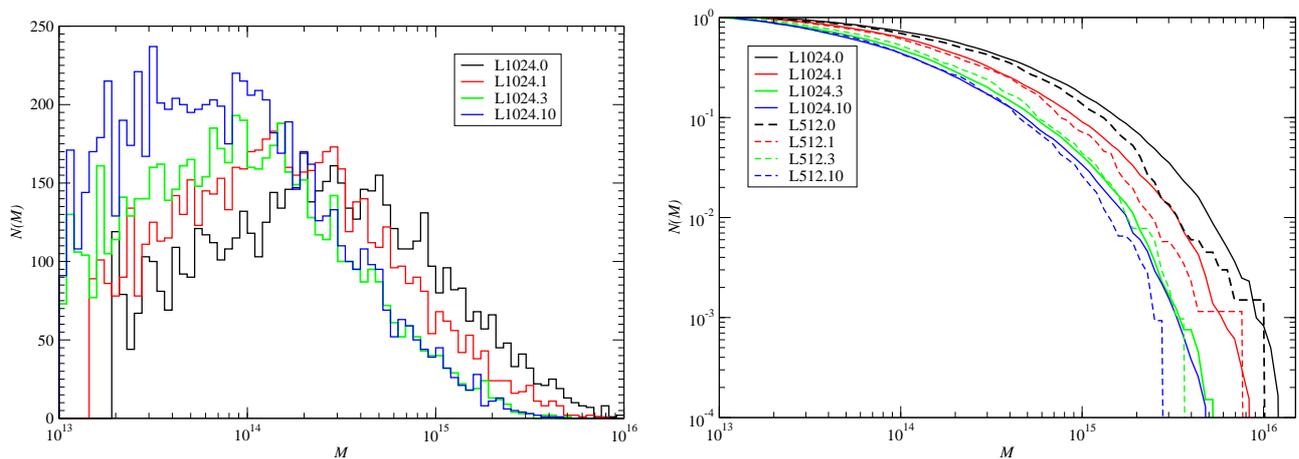
 
\centering 
\hspace{2mm}
\resizebox{0.45\textwidth}{!}{\includegraphics*{L1024_scl2_lumdistrTnew.eps}}
\hspace{2mm}
\resizebox{0.45\textwidth}{!}{\includegraphics*{L1024_512_scl2_lumdistrT.eps}}
\caption{{ Evolution of supercluster masses of LCDM models.}
  {\em Left:} Distribution of supercluster masses of the
  LCDM model at various epochs. {\em Right:} Cumulative distributions
  of supercluster masses of the LCDM models. For comparison we show  the
  cumulative supercluster mass distribution for a $\Lambda$CDM model
  with side length $L_0=512$~\Mpc. Masses are given in solar
  units.  }
\label{fig:Fig10} 
\end{figure*}

\subsection{Comparison with superclusters found from velocity data}

The property of superclusters to act as great attractors was the basis
of the suggestion of \citet{Tully:2014}, \citet{Pomarede:2015aa} and
\citet{Graziani:2019aa} that superclusters should be defined on the
basis of their dynamical effect on the cosmic environment, the
so-called basins of attraction (BoA).  To keep the term
``superclusters'' in its 
conventional meaning, \citet{Einasto:2019fk} suggested to use the term
``cocoons'' for Tully's BoAs.  Neighbouring cocoons have common
sidewalls.  \citet{Einasto:2019fk} suggested that a good measure for
estimating the size of cocoons is the fitness diameter, which remains
almost constant during the cosmic evolution in co-moving coordinates.
According to the definition of the cocoon sizes through the fitness
diameter, cocoons  fill the whole volume of the
observable universe, see Eq.~(\ref{fitness}).

\citet{Dupuy:2019aa} applied constrained simulations in a $\Lambda$CDM
model of length 500~\Mpc, based on data from Cosmicflows-2 \citep{Tully:2013aa}
and Cosmicflows-3 \citep{Tully:2016aa}, and identified several
BoAs.  The BoA diameters calculated from volumes $D = V^{1/3}$
are 79, 89, and 100~\Mpc\ for the Laniakeia, Perseus-Pisces, and Coma
supercluster BoAs, respectively.  According to our analysis, the SDSS
sample has 844 superclusters. The fitness diameter of the largest SDSS
supercluster cocoon is 132~\Mpc, the median fitness diameter of
the cocoons is 41~\Mpc, and the fitness diameters of 10~\% of largest cocoons
are larger than or equal to 84~\Mpc.  The methods for defining the sizes of BoAs and
our cocoons are different (velocity and density fields, respectively),
but the numerical results for the sizes are similar.

\citet{Dupuy:2020aa} used the SmallMultiDark simulations by
\citet{Klypin:2016aa} to segment the universe into dynamically
coherent basins applying various smoothing lengths to velocity
data. This simulation was performed in a box of size 400~\Mpc, using
$3840^3$ particles. The density and velocity fields were calculated in a
$256^3$ grid. The evolution of basins was followed in redshift intervals
from $z=2.89$ to $z=0$. To test the effect of smoothing, the final
velocity field was Gaussian smoothed with dispersions $r_s=1.5$ to
15~\Mpc. Basins were searched for using three parameters: 
the smoothing length, the maximum streamlines length, and the
integration step along streamlines.  At optimal search parameters, the
number of basins converged to 647 for a smoothing scale 1.5~\Mpc, to
about 250 basins for a smoothing scale 3~\Mpc, and to only a few for
smoothing scale 15~\Mpc.  Taking the size of the
simulation box into account, these numbers agree fairly well with the
number of SDSS superclusters and with the respective number in our present
LCDM model. As in our LCDM model, the number of the \citet{Dupuy:2020aa}
basins decreases slightly with cosmic epoch from about 760 at
$z=2.89$ to 647 at $z=0$.  The mass distribution of the basins is
almost independent of the redshifts, see Figs.~9 and 10 of
\citet{Dupuy:2020aa}.

It is interesting to compare the evolution of supercluster cocoons and
superclusters.  We show in Fig.~\ref{fig:Fig10} the distribution of
supercluster masses in our LCDM model at various epochs. The left
panel of Fig.~\ref{fig:Fig10} shows the distribution of supercluster
masses, and the right panel shows the cumulative distribution of
supercluster masses, normalised to the total number of superclusters.
For comparison we also show the cumulative distribution of
supercluster masses for an LCDM model of size $L_0=512$~\Mpc\ by
\citet{Einasto:2019fk}.  { The comparison of the evolution of
  superclusters and cocoons shows that number of superclusters as well
  cocoons remains almost constant during the evolution. } The
differences in the evolution are more interesting, however.  The
cocoon masses remain constant during the evolution
\citep{Dupuy:2020aa}, whereas the supercluster masses increase during
the evolution by a factor of about 3, see the right panel of
Fig.~\ref{fig:Fig10}.

The second important difference is in masses themselves.
Fig. ~\ref{fig:Fig10} shows that the most massive LCDM superclusters
have masses $M \approx 10^{16} M_{\odot}$ at the present epoch.
Fig.~9 of \citet{Dupuy:2020aa} shows that most massive supercluster
basins (cocoons) have masses $M \approx 2 \times 10^{17} M_{\odot}$ at
all epochs.  This difference is expected.  The supercluster volumes at
the early epoch are about 50 times smaller than the supercluster
cocoon volumes; at the present epoch, this difference has increased to
about 140 times, see Eq.~(\ref{fitness}), Table~\ref{Tab2}, and
Fig.~\ref{fig:Fig8}.  {  The difference of  masses of  superclusters  and cocoons
at the present epoch is only by a factor of about 20. } This means that cocoon regions
outside superclusters have much lower densities than inside
superclusters. Because the cocoon masses remain constant during the
evolution, the growth of supercluster masses can be explained by the
infall of surrounding matter inside cocoons to superclusters.{ To
  illustrate the growth of superclusters, we show in the upper and
  middle panels of Fig.~\ref{fig:Mdenfield} the LCDM model density
  fields at redshifts $z=30,~10,~3,\text{and}~0$ at two smoothing
  scales, 2~\Mpc\ and 8~\Mpc. The supercluster contraction can be
  followed in the upper panels, most strongly at redshift $z<1$.  See
  also Fig.~\ref{fig:Fig10} for the increase in supercluster masses at
  $z<1$.}  The exchange of matter between cocoons is minimum because
the velocity flow within the cocoons is directed inwards.

Fig.~\ref{fig:Fig10} shows that the mass distribution of
superclusters in models of size 512 and 1024~\Mpc\ is approximately
similar.  However, the model of size 1024~\Mpc\ contains slightly more
massive superclusters at all simulation epochs.  This small
difference can be explained by the larger volume of the 1024~\Mpc\
model: there are 6044 superclusters in the 1024~\Mpc\  model versus 995
superclusters in the L512 model.

\subsection{Evolution of superclusters and supercluster cocoons}

{ Available data suggest that galaxy and supercluster embryos  were
  created by high peaks in the initial field.  The initial 
  velocity field around peaks is almost laminar.  The development of
  the density field in the early phase is well described by the 
  \citet{Zeldovich:1970} approximation (ZA) and its extension, the
  adhesion model by \citet{Kofman:1988aa}.  As shown by
  \citet{Kofman:1990aa, Kofman:1992aa}, the adhesion approximation
  yields structures that are very similar to structures calculated with N-body
  numerical simulations of the cosmic web evolution with the
  same initial fluctuations.  \citet{Bernardeau:1995aa} compared the
  evolution of density fields  by ZA with the results obtained by
  various other approximations, and found that  various methods agreed well.

  In the modern simulations of the cosmic web evolution we applied
  here, the early evolution was calculated using ZA. The 
  structure at the first epoch used in our study, corresponding to 
  redshift $z=30$, is obtained by ZA.    Fig.~\ref{fig:Mdenfield}
  shows that the density distributions at epochs $z=30$ and 
  $z=10$ are very similar, only the fluctuation amplitude has
  increased.  Large-scale supercluster-type structures are clearly visible
  at all epochs, and the growth of superclusters inside cocoons is
  well visible.  On supercluster
  scales the results obtained for different epochs and
  simulation cube sizes agree very well; see  \citet{Einasto:2020ab}.  }

\citet{Pichon:2011aa} and \citet{Dubois:2012aa, Dubois:2014aa} showed
that a significant fraction of the cold gas falls along filaments
that are oriented nearly radially to the centres of high-redshift rare massive
haloes. This process rapidly increases the mass of the central
halo. We may conclude that depending on the height of the initial 
density peak, galaxy embryos, galaxy clusters,
and superclusters were created in this way. However, the further evolution of
superclusters differs from the evolution of galaxies and galaxy
clusters.  Galaxies and ordinary galaxy clusters are local 
attractors and collect additional matter from their local
environment. Supercluster are global attractors and collect matter
from a much larger environment.

The density field method applied in this paper allows us to select
superclusters of the cosmic web and to estimate the size of their
cocoons.  As discussed above, at the early epoch, superclusters are
approximately 50 times smaller than their cocoons, and at the present
epoch, they are about 140 times smaller.  { Our density field method
  allows us to select individual superclusters and to find statistical
  properties of the ensemble of cocoons.  The velocity field method
  allows us to find individual supercluster cocoons, but not
  superclusters themselves.} Thus these methods are complementary.

The filamentary character of the cosmic web can be described using the
skeleton, that is, the 3D analogue of ridges in a mountainous landscape
\citep{Pichon:2010aa}.  Peaks of the cosmic web are connected by
filaments. The number of filaments that connect the clusters with other
clusters can be called the connectivity for global connections (including
bifurcation points), and multiplicity for local connections
\citep{Codis:2018aa}.  \citet{Kraljic:2020aa} investigated the
connectivity of the SDSS galaxy sample. The authors first determined
the skeleton of the SDSS sample, traced by the DisPerSE algorithm by
\citet{Sousbie:2011aa}.  Then they calculated the connectivity of
all clusters. They found  that the connectivity of the
SDSS sample clusters has a peak at 3, and the multiplicity (local connectivity)
has a peak at 2.  Both parameters depend on the cluster mass. The mean
connectivity  of massive SDSS clusters is 4,  and the 
multiplicity of most massive clusters is 6. 

These results have a simple explanation. Low- and medium-mass clusters
lie inside filaments, and thus have the multiplicity 2 (the connection is
from both sides of the cluster inside the filament).  Clusters move
together with  filaments in the large potential well of
superclusters. The simultaneous movement of clusters with their
surrounding filament follows from the simple fact that the filamentary
character of the cosmic web is preserved at the present epoch. If
clusters had high peculiar velocities with respect to
the surrounding filaments, the filamentary
character of the web would be destroyed during the evolution. The laminar character of the
velocity field is explained by the presence of the DE, as
suggested already by \citet{Sandage:2010aa}.  Very rich clusters are
central clusters of superclusters, and they are connected with other
structures by many filaments. This has been demonstrated by
\citet{Tully:1978} for the Virgo supercluster, by
\citet{Joeveer:1977lj}, \citet{Joeveer:1978dz} and \citet{Joeveer:1978pb} for
the Perseus-Pisces supercluster, and by \citet{Einasto:2020cc} for the
A2142 supercluster.  The central clusters of these superclusters lie at
minima of potential wells created by respective superclusters.  They
are fed by filaments from several sides, and are suitable locations
for cluster merging, which means that small clusters fall onto the central cluster
along filaments surrounding the central cluster. The pattern of the
cosmic web suggests that the high connectivity can be used as a
signature for the presence of a central cluster of a supercluster.

\section{Summary remarks}

We calculated the percolation functions of superclusters for four
evolutionary epochs of the Universe, corresponding to redshifts
$z=30, ~10,~3,~1$, and  $z=0$. The analysis was made
for four sets of  cosmological models: the LCDM model, the classical
standard SCDM model, the open OCDM model, and the hyper-DE 
HCDM model. Ensembles of superclusters were found for these four
models for all evolutionary epochs. All models have the same initial
phase realisation, so that we can follow  the role of different 
values of cosmological parameters in the evolution of superclusters
and their cocoons. 

The almost constant number of superclusters and of the
{  (co-moving)}  volume of 
cocoons during the evolution means that supercluster embryos
were created at a very early evolution epoch, much earlier than epochs
tested in this study, $z=30$. On the other hand, the
differences in numbers and volumes between different cosmological
models at all epochs suggest that these differences were also created  at an early
epoch, most likely after the end of inflation and before
matter and radiation equilibrium. The differences between
models suggest that  the supercluster and supercluster  cocoon properties
as measured by the percolation method using the density field and by
the velocity field as done by \citet{Dupuy:2020aa} can be used to test the basic
cosmological model parameters.

We analysed the evolution of superclusters and their cocoons by applying
percolation functions of the density field.  \citet{Dupuy:2020aa}
studied the evolution of supercluster basins of attraction using the
velocity field.  Both methods yield an approximately equal spatial
density of supercluster cocoons with rather similar properties. This
similarity suggests that the density as well the velocity fields can
be used to detect superclusters and their cocoons and to
investigate their properties and evolution. The velocity field is
physically more justified when velocity data are available. It can only
be used today to study the nearby space of the real universe, however. For
more distant regions the density field method is 
the only option, at least at the moment.  The velocity-field-based method allows us to separate
individual supercluster cocoons, and the density field method allows us to
separate individual superclusters.

A more detailed study of  the differences in the evolution of  the LCDM,
OCDM, and HCDM models is beyond the scope of the present paper.
The basic conclusions of our study are listed below.
 
\begin{enumerate} 

\item{} The combined analysis of the density and velocity field
  evolution shows that superclusters and 
their cocoons evolve differently. 
  
\item{} Volumes (in comoving coordinates), masses, and numbers of
  supercluster cocoons are stable parameters and are almost identical
  for all evolutionary epochs.  This suggests that embryos of
  supercluster cocoons were created at an early epoch. At epoch
  $z=30,$ superclusters have volumes about 50 times lower than their
  cocoons. At the present epoch, supercluster volumes are about 140
  times lower than the volumes of their cocoons. Supercluster masses
  are about 20 times lower than the cocoon masses.  Supercluster
  masses increase by about a factor of 3 during the evolution, and
  their volumes (in co-moving coordinates) decrease by about the same
  factor. Superclusters mainly evolve inside their cocoons.

\item{} The LCDM, OCDM, and HCDM models have almost similar
  percolation parameters.  This suggests that the essential parameter
  that defines the supercluster evolution is the matter density.  The
  DE density affects the growth of the density perturbation amplitude
  and the growth of supercluster masses, albeit significantly weaker.
  The HCDM model has the highest growth speed of the density
  fluctuation amplitude and the largest growth of supercluster masses
  during the evolution.  The geometrical diameters and numbers of HCDM
  superclusters at high threshold densities are larger than for LCDM
  and OCDM superclusters.  The SCDM model has about twice more
  superclusters than the other models. The SCDM superclusters are
  smaller, and their mass is lower than in other models.

\end{enumerate}

\begin{acknowledgements}
  
Authors thank the anonymous referee for useful suggestions.
  This work was supported by institutional research funding IUT40-2 of
  the Estonian Ministry of Education and Research, by the Estonian
  Research Council grant PRG803, and by Mobilitas Plus grant MOBTT5. We
  acknowledge the support by the Centre of Excellence``Dark side of
  the Universe'' (TK133) financed by the European Union through the
  European Regional Development Fund.

  We thank the SDSS Team for the publicly available data releases. 
  Funding for the SDSS and SDSS-II has been provided by the Alfred 
  P. Sloan Foundation, the Participating Institutions, the National 
  Science Foundation, the U.S. Department of Energy, the National 
  Aeronautics and Space Administration, the Japanese Monbukagakusho, 
  the Max Planck Society, and the Higher Education Funding Council for 
  England. The SDSS Web Site is \texttt{http://www.sdss.org/}.

The SDSS is managed by the Astrophysical Research Consortium for the 
Participating Institutions. The Participating Institutions are the 
American Museum of Natural History, Astrophysical Institute Potsdam, 
University of Basel, University of Cambridge, Case Western Reserve 
University, University of Chicago, Drexel University, Fermilab, the 
Institute for Advanced Study, the Japan Participation Group, Johns 
Hopkins University, the Joint Institute for Nuclear Astrophysics, the 
Kavli Institute for Particle Astrophysics and Cosmology, the Korean 
Scientist Group, the Chinese Academy of Sciences (LAMOST), Los Alamos 
National Laboratory, the Max-Planck-Institute for Astronomy (MPIA), 
the Max-Planck-Institute for Astrophysics (MPA), New Mexico State 
University, Ohio State University, University of Pittsburgh, 
University of Portsmouth, Princeton University, the United States 
Naval Observatory, and the University of Washington.

\end{acknowledgements}

\bibliographystyle{aa} 

\end{document}